\documentclass[letterpaper]{jpconf}
\usepackage{graphicx,color}
\usepackage{natbib}

\newcommand{\be}{\begin{eqnarray}}
\newcommand{\ee}{\end{eqnarray}}

\newcommand\simgreater{\,\lower0.7ex\hbox{$\stackrel{>}{\sim}$}\,}
\newcommand\simless{\,\lower0.7ex\hbox{$\stackrel{<}{\sim}$}\,}
\newcommand\msol{$M_\odot$}

\newcommand{\nue}{\ensuremath{\nu_{{\rm e}}}}
\newcommand{\nuebar}{\ensuremath{\bar{\nu}_{{\rm e}}}}

\newcommand{\nux}{\ensuremath{\nu_{{\rm x}}}}
\newcommand{\nuxbar}{\ensuremath{\bar{\nu}_{{\rm x}}}}
\newcommand{\num}{\ensuremath{\nu_{\mu}}}
\newcommand{\numbar}{\ensuremath{\bar{\nu}_{\mu}}}
\newcommand{\nut}{\ensuremath{\nu_{\tau}}}
\newcommand{\nutbar}{\ensuremath{\bar{\nu}_{\tau}}}

\begin{document}

\title{2D and 3D core-collapse supernovae simulation results obtained with the CHIMERA code}

\author{S. W. Bruenn$^{1}$, A. Mezzacappa$^{2}$, W. R. Hix$^{2}$, J. M. Blondin$^{3}$, P.  Marronetti$^{1}$, O. E. B. Messer$^{4}$, C. J. Dirk$^{1}$ and S. Yoshida$^{5}$}

\address{$^{1}$ Physics Department, Florida Atlantic University, 777 W. Glades Road, Boca Raton, FL 33431-0991}

\address{$^{2}$ Physics Division, Oak Ridge National Laboratory, Oak Ridge, TN 37831--6354}

\address{$^{3}$ Department of Physics, North Carolina State University, Raleigh, NC 27695-8202}

\address{$^{4}$ Center for Computational Sciences, Oak Ridge National Laboratory, Oak Ridge, TN 37831--6354}

\address{$^{5}$ Max-Planck-Institut fur Gravitationsphysik, Albert Einstein Institut, Golm, Germany}

\ead{bruenn@fau.edu}

\begin{abstract}
Much progress in realistic modeling of core-collapse supernovae has occurred recently through the availability of multi-teraflop machines and the increasing sophistication of supernova codes. These improvements are enabling simulations with enough realism that the explosion mechanism, long a mystery, may soon be delineated. We briefly describe the CHIMERA code, a supernova code we have developed to simulate core-collapse supernovae in 1, 2, and 3 spatial dimensions. We then describe the results of an ongoing suite of 2D simulations initiated from a 12, 15, 20, and 25 M$_{\odot}$ progenitor. These have all exhibited explosions and are currently in the expanding phase with the shock at between 5,000 and 20,000 km. We also briefly describe an ongoing simulation in 3 spatial dimensions initiated from the 15 M$_{\odot}$ progenitor.
\end{abstract}

\section{Introduction}
\label{sec:Intro}

The neutrino transport mechanism is still the leading explosion mechanism candidate for core-collapse supernova arising from stars with initial mass of $\sim 8 - 25$ \msol. Originally proposed by \citep{colgate_w66}, it asserts that the fraction of neutrinos absorbed by the outer layers that are released by the inner core during and following its collapse deposit sufficient energy to power the explosion. For many years this mechanism was investigated by computer modeling with the assumption of spherical symmetry, and by the turn of this century these simulations had achieved considerable realism with codes  employing Boltzmann neutrino transport, state-of-the-art neutrino interactions, and general relativity \citep{mezzacappa_lmhtb01, liebendorfer_mtmhb01, liebendorfer_rjm05}. These simulations found that the shock wave, launched by the bounce of the core, stalled at a radius between $\sim$ 100 - 200 km rather than giving rise to an explosion.

While the spherically symmetric simulations failed to explode, other investigations taking place during this period of time pointed to the importance of multi-dimensional effects. In particular, a variety of fluid instabilities are present or quickly develop in the immediate post-bounce core \citep{bethe_90}. The most important of these for the neutrino heating mechanism is neutrino-driven convection---convection driven by the neutrino heating above the neutrinosphere. Because neutrinos heat the bottom of the heating layer most intensely a negative entropy gradient builds up which renders the layer convectively unstable. For convection to grow, however, the fluid must remain in the heating layer for a critical length of  time. In particular, the ratio of the advective timescale, $\tau_{\rm adv}$, to some averaged timescale of convective growth, $\tau_{\rm cv}$, must be $\stackrel{\textstyle>}{\sim}$ 3 \citep{foglizzo_sj06}. If convection gets established in the heating layer, hot gas from the neutrino-heating region will be transported directly to the shock, while downflows simultaneously will carry cold, accreted matter to the layer of strongest neutrino heating where a part of this gas, being cold, readily absorbs more energy from the neutrinos. The loss of energy accompanying the advection of matter through the gain radius is thereby reduced and more energy stays in the heating layer. 

Attaining the magic value 3 of $\tau_{\rm adv}/\tau_{\rm cv}$ is aided by the instability of the stalled shock to low-mode aspherical oscillations, which are referred to as the standing accretion shock instability or `SASI.' Discovered by \citep{blondin_md03}, the development of the SASI leads to an enlargement of the heating layer in one region and its diminution in another. This results in an enhanced efficiency of neutrino heating and more favorable conditions for convection in the enlarged region, and conditions more favorable for the establishment of down-flows or return-flows in the constricted region. In addition to promoting an explosion, neutrino-driven convection and the SASI can lead to the large asymmetries observed for SN 1987A and other supernovae, and might account for the large observed velocities of neutron stars \citep{kifonidis_psjm06, scheck_kjm06}.

\section{The CHIMERA Code}

The CHIMERA code is designed to simulate core-collapse supernovae in 1, 2, and 3 spatial dimensions and with realistic spectral neutrino transport from the onset of collapse to the order of 1 sec post bounce with the present generation of supercomputers. It conserves total energy (gravitational, internal, kinetic, and neutrino) to within $\pm$ 0.5 B, given a conservative gravitational potential. The code currently has three main components:  a hydro component, a neutrino transport component, and a nuclear reaction network component. In addition there is a Poisson solver for the gravitational potential and a sophisticated equation of state. A preliminary version of the code was briefly described in \cite{bruenn_dmhbhm06, mezzacappa_bbhb07, bruenn_mhbmmdy09} and we briefly highlight the main features here.

The hydrodynamics is performed via a Lagrangian remap implementation of the  Piecewise Parabolic Method (PPM) \citep{colella_w84}. A moving radial grid option wherein the radial grid follows the average radial motion of the fluid makes it possible for the core infall phase to be followed with good resolution. Following bounce, an adaptive mesh redistribution algorithm keeps the radial grid between the core center and the shock structured so as to maintain approximately constant $\Delta \rho/\rho$. For 256 and 512 radial zones, this ensures that there are at least 15 and 30 radial zones per decade in density, respectively. The equation of state (EOS) is standard and is described in \citep{bruenn_dmhbhm06}. The algorithm described in \citep{sutherland_bb03}, modified for greater robustness, was employed to stabilize shocks oriented along grid lines. Gravity is computed by an approximate general relativistic potential for spherical gravity and Newtonian 2D/3D spectral Poisson solver for the higher moments..

Neutrino transport is implemented by a ``ray-by-ray-plus'' approximation \citep[cf.][]{buras_rjk06a} whereby the lateral effects of neutrinos such as lateral pressure gradients (in optically thick conditions), neutrino advection, and velocity corrections are taken into account, but transport is performed only in the radial direction. Transport is computed by means of multigroup flux-limited diffusion with a sophisticated flux limiter that has been tuned to reproduce results of a general relativistic Boltzmann transport solver to within a few percent \citep{liebendorfer_mbmbct04}. All O($v/c$) observer corrections have been included. The transport solver is fully implicit and solves for four neutrino flavors simultaneously (i.e., \nue's, \nuebar's, \num's and \nut's (collectively \nux's), and \numbar's and \nutbar's (collectively \nuxbar's)), allowing for neutrino-neutrino scattering and pair-exchange, and different $\nu$ and $\bar{\nu}$ opacities. State-of-the-art neutrino interactions are included with full energy dependences.

The nuclear composition in the non-NSE regions of these models is evolved by the thermonuclear reaction network of \citep{hix_t99a}.  This is a fully implicit general purpose reaction network, however in these models only reactions linking the 14 alpha nuclei from $^{4}$He to $^{60}$Zn are used. Data for these reactions is drawn from the REACLIB compilations \citep{rauscher_t00}.

\section{Simulation Results}

\begin{figure}[!h]
\vspace*{0cm}
\setlength{\unitlength}{1.0cm}
\begin{minipage}[t]{2.9in}
\hspace*{-1.cm}
\vspace*{0cm}
{\includegraphics[angle=-90,scale = 0.39,viewport = 700 10 20 20]{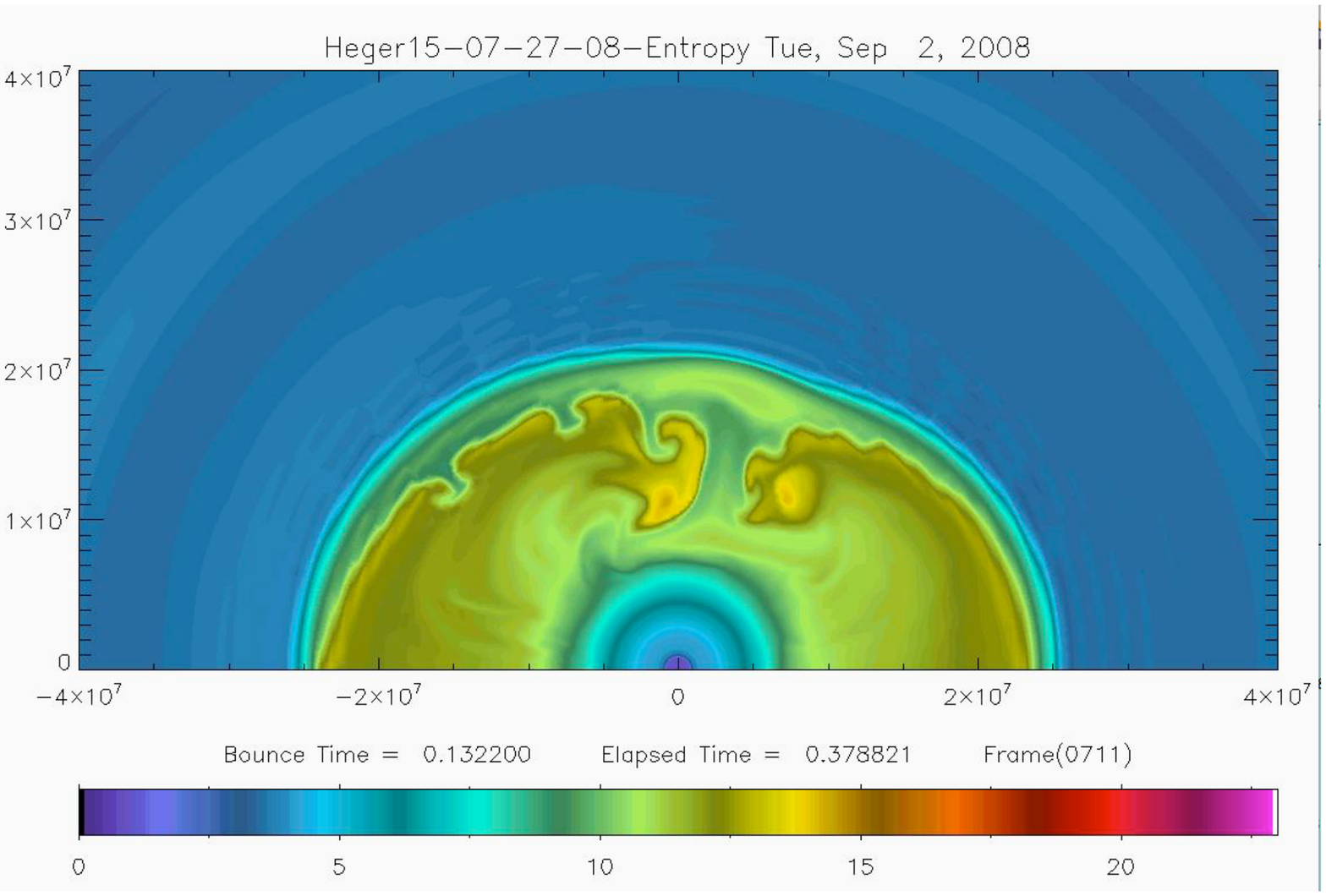}}
\caption{\label{fig:bwfig_1}
\small{Snapshot of the entropy distribution of the 15 M$_{\odot}$ 2D simulation at 132 ms post-bounce time.}}
\end{minipage}
\hfill
\begin{minipage}[t]{2.9in}
\hspace*{-0.7cm}
{\includegraphics[scale = 0.25]{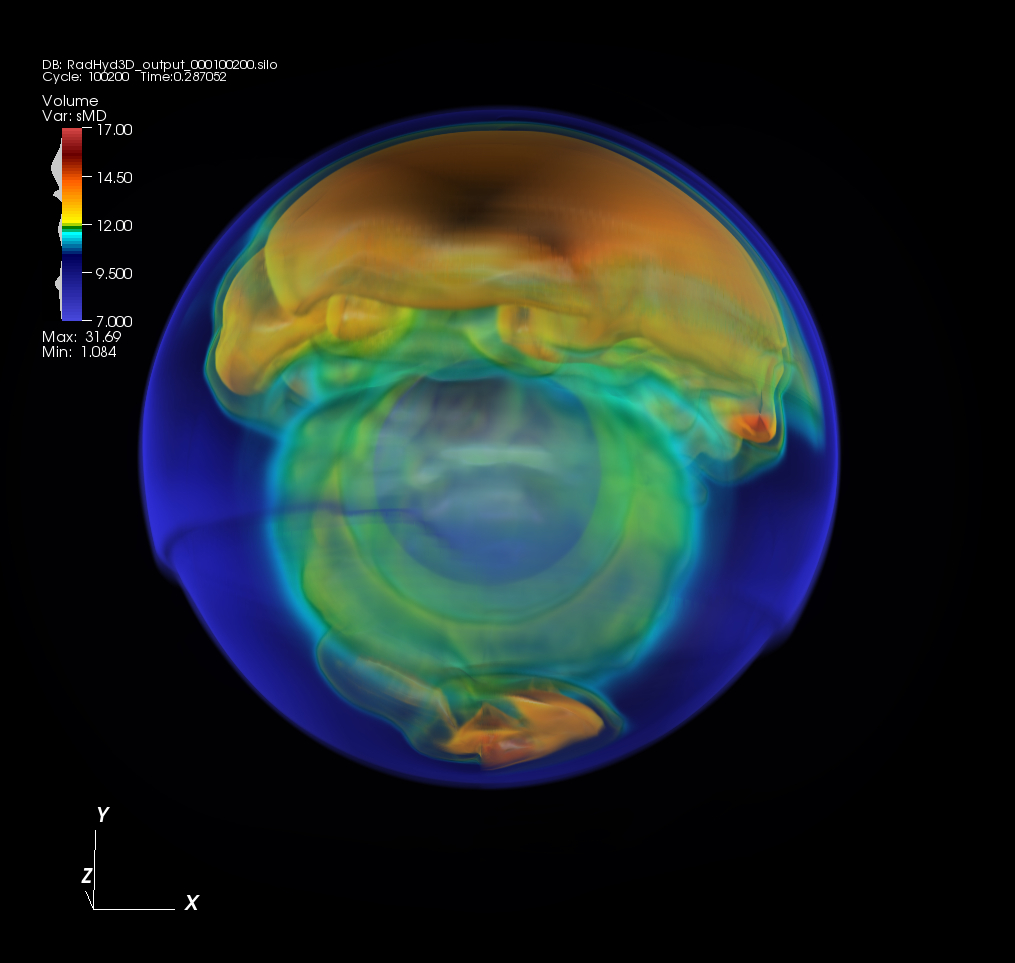}}
\vspace*{-0.5cm}
\caption{\label{fig:bwfig_2}
\small{Snapshot of the entropy distribution of the 15 M$_{\odot}$ 3D simulation at 132 ms post-bounce time.}}
\end{minipage}
\end{figure}

Simulations were performed for 12, 15, 20, and 25 \msol\ progenitors \citep{woosley_h07}. A resolution of 256x256 radial and angular zones was used. The simulations were repeated for the 15 \msol\ progenitor with resolutions of 512x128 and 512x256. With the many improvements in the CHIMERA code and additions of more realistic physics (e.g., GR corrections, state-of-the-art neutrino microphysics) since our first results were reported in \citep{bruenn_dmhbhm06} we find that all our models explode as before, but now improved neutrino microphysics rather than nuclear reactions play a key role. For a period of time after bounce (65 ms for the 12 and 15 M$_{\odot}$ models, 105 ms and 130 ms respectively for the 20 and 25 M$_{\odot}$ models) the 2D and corresponding spherically symmetric simulations track each other very closely. During this period the heating layer becomes convectively unstable but its growth is suppressed in the 2D simulations because of the too rapid inflow of material. At the end of this period $\tau_{\rm adv}/\tau_{\rm cv}$ exceeds 3 in some regions and convective mushrooms begin to appear in the heating layer and at almost the same time the shock begins to exhibit SASI dipole and quadrupole deformations along the polar axis. This expands the heating layer in the polar regions and the enhanced convection in these regions quickly develops into a large-scale overturn (Fig. \ref{fig:bwfig_1}), with material rising and expanding in the polar regions and flowing towards the equatorial region. By 225 ms post-bounce for the 15 M$_{\odot}$, as an example, large lobes of high-entropy neutrino heated gas are pushing the shock outwards in the polar regions while an equatorial accretion funnel has developed which is channeling lower-entropy newly shocked material down to the gain layer. The narrow end of the accretion funnel tends to oscillate slowly from one hemisphere to the other, seeming thereby to alternately pump up one hemisphere and then the other with energy by the enhanced neutrino emission at its base. 

\begin{figure}[!h]
\vspace{0.cm}
\hspace{2.5cm}
\setlength{\unitlength}{1.0cm}
{\includegraphics[width=11.0 cm]{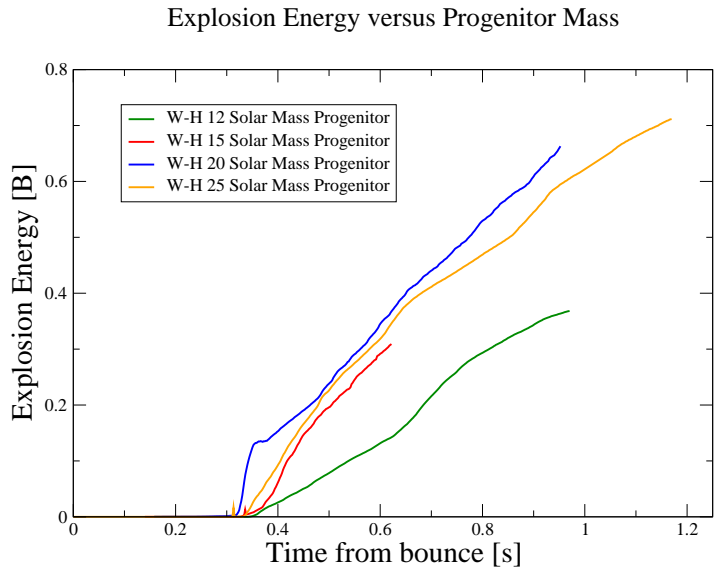}}
\caption{\label{Explosion}
Explosion energies as a function of post-bounce time.}
\end{figure}

Eventually a runaway condition is met and the shock begins to accelerate rapidly outwards. Measured by the first signs of positive radial post-shock velocities (a different criterion than that used by the Garching group) explosions commence at roughly 300 ms from bounce for the four models. The 12, 15, and 25 M$_{\odot}$ models exhibit a pronounced quadrupolar prolate configuration of the shock as the explosion commences, while the 20 M$_{\odot}$ the configuration of the shock is more dipolar. The ratio of the major to minor axis of the shock ranges from 1.5 (25 M$_{\odot}$ model) to 2.0 (15 M$_{\odot}$ model). The explosion energies, computed conservatively as the sum of the kinetic, internal thermal, and kinetic energies of outward moving material is shown in Figure \ref{Explosion}. These explosion energies are clearly asymptoting to observed supernovae energies. 

A medium resolution simulation in 3 spatial dimensions with 304 radial zones, 76 angular, and 152 azimuthal zones initiated from the 15 M$_{\odot}$ progenitor is also in progress. Figure (\ref{fig:bwfig_2}) shows a the core configuration at the same post-bounce time as the corresponding 2D simulation shown in Figure (\ref{fig:bwfig_1}). The 3D simulation is clearly exhibiting large scale convection patterns emerging from the poles at this time. It's further evolution should validate or throw into question conclusions drawn from 2D simulations.

\ack
The authors would like to acknowledge partial funded by the U.ÊS. Department of Energy under Contract No. DEAC05-00OR22725 at Oak Ridge National Laboratory, the resources of the National Center for Computational Sciences at Oak Ridge National Laboratory, and a TACC Ranger (TG-MCA08X010) computational award. S. W. B., P. M. and O. B. E. M. acknowledge partial support from an NSF-OCI-0749204 award, S. W. B., P. M., O. B. E. M., W. R. H., and A. M. acknowledge partial support from NASA award (07-ATFP07-0011). A.M., W.R.H., and O.E.B.M. are supported at the Oak Ridge National Laboratory, managed by UT-Battelle, LLC, for the U.S. Department of Energy under contract DE-AC05-00OR22725.

\bibliography{BibTeX_list}

\end{document}